# Comparison of Casimir forces and electrostatics from conductive SiC-Si/C and Ru surfaces

Z. Babamahdi[1], V. B. Svetovoy[1,2], M. Enache[1], M. Stöhr[1], and G. Palasantzas[1]

[1] Zernike Institute for Advanced Materials, University of Groningen, Nijenborgh 4, 9747 AG Groningen, The Netherlands

[2] A. N. Frumkin Institute of Physical Chemistry and Electrochemistry, Russian Academy of Sciences, Leninsky prospect 31 bld. 4, 119071 Moscow, Russia

**Abstract**

Comprehensive knowledge of Casimir forces and associated electrostatics from conductive SiC and Ru surfaces can be essential in diverse areas ranging from micro/nanodevice operation in harsh environments to multilayer coatings in advanced lithography technologies. Hence, the Casimir force was measured between an Au-coated microsphere and N-doped SiC samples with Si- and C-terminated faces, and the results were compared with the measurements using the same microsphere and a metallic Ruthenium surface. Electrostatic calibration showed that the Si- and C-faces behave differently with a nearly ~0.6-0.7 V difference in the contact potentials $V_0^{Si/C}$. We attribute this to a higher incorporation of N on the C-terminated face in the near surface region resulting in the formation of $NO_x$ and an increased work function compared to the Si-terminated surface which is in agreement with x-ray photoelectron spectroscopy data. Notably, the contact potential of the SiC-C face ($V_0^C$ ~ 0.1 V) was closer to the metallic Ru-Au system ($V_0^{Ru}$ ~0.05 V).

However, the measured optical properties of the SiC-Si/C terminated surfaces with ellipsometry did not show any substantial differences indicating that the effective depth of the Si/C terminating surface layers are significantly smaller than the photon penetration depth not leading to any differences in the calculated forces via Lifshitz theory. Nonetheless, the measured Casimir forces, after compensation of the electrostatics contributions, showed differences between the Si/C faces, whereas the comparison with the Lifshitz theory prediction shows better agreement for the SiC-Si face. Finally, comparison of the Casimir forces below 40 nm separations between the SiC-Si/C and Ru surfaces indicated that the short-range roughness effects on the Casimir force increase in magnitude with increasing metallic behavior of the plate surface. Therefore, not only the material optical properties but also the conductive state and roughness of the surface layers must be carefully taken into account in short range Casimir interactions between more complex dielectric materials.

# I. Introduction

Nowadays, the Casimir force that originates from the perturbation of electromagnetic vacuum fluctuations is still a topic of relentless research [1-21], though its proposition in 1948 by the Dutch physicist Hendrik Casimir almost dates back 70 years [1]. The interest stems from a multitude of research fields ranging from fundamental physics in search of new forces beyond the standard model to micro/nanodevices for technology applications [2–6]. Lifshitz and co-workers in the 1950s [7,8] considered the general case of flat dielectric plates by exploiting the fluctuation-dissipation theorem, which relates the dissipative properties of the plates (due to optical absorption by many microscopic dipoles) and the resulting electromagnetic (EM) fluctuations. The theory describes the attractive interaction due to quantum fluctuations for all separations covering both the Casimir (long-range) and van der Waals (short-range) regimes [2–9]. As devices enter the submicron range within the realm of nanoelectromechanical systems, a deep understanding of Casimir forces at nanoscale separations between real materials is inevitable.

The dependence of the Casimir force on the type of material is an important topic since in principle one can tailor the force by a suitable choice of materials [6,8,18–21,10–17]. In this sense, silicon-based semiconductors appear as a promising choice since the materials properties can be modified and controlled allowing also the tuning of the Casimir interactions [6,21]. Different doping and charge carrier densities in semiconductors can change the Casimir force [10]. However, in microelectromechanical systems applications for industry, e.g. automotive and space technologies [22–24], the micro-sensors are required to operate in harsh environments. Because the latter can be a challenge for Si-based sensing devices, silicon carbide (SiC) is considered an excellent substitute for Si due to its outstanding properties if attributes such as high durability combined with high stiffness and low thermal expansion are necessary. SiC is currently also utilized for precise instrumentation frames and mirrors, and additionally, there is the possibility for usage in macro/nano assembly technologies via direct (optical) bonding concepts [25–27]. In addition, because SiC exhibits high hardness, chemical inertness, and ability to survive operation at high temperatures and harsh environments, it is well suited to be used as protective coating of micro-machined parts.

SiC is a material that exhibits strong polytypism. All polytypes have identical planar arrangement, while their difference lies in the stacking of the planes so that the difference in the stacking periodicity of similar planes results in different types of polytypes (≥250). The relatively low residual stress level in the layers, the high stiffness, and the excellent etch-stop properties allow even the fabrication of free standing SiC microstructures via standard Si bulk micro-machining techniques [25–27]. The terminating layers of SiC can be either Si or C and that might alter the associated Casimir interaction for this material in potential applications, taking also the associated electrostatic characteristics (e.g. contact potentials) into consideration. Although Casimir force measurements on Si-terminated nitrogen doped SiC have been performed in the sphere-plate geometry, which gave reasonable agreement with predictions of the Lifshitz theory, [28] the effect of the surface termination layer has remained so far unexplored.

Therefore, measurements and analysis of the Casimir forces at short ranges (< 100 nm) from both Si/C faces are necessary to gauge the effect of the surface termination on both the Casimir force and the associated surface electrostatics. In order to minimize charging and electrostatic effects, since SiC is insulating, our measurements were performed on highly doped conductive SiC samples. Moreover, any roughness on the termination layers of the SiC and/or on the colloid probe can have significant influence on the Casimir forces, as studies for metal coatings [29–34] have shown, whereas its magnitude in relation to the conductivity of a dielectric system still remains unexplored. In fact, increasing the minimum separation between interacting surfaces from zero, for an ideal smooth system, to $d_0$ (distance upon contact due to roughness) for a rough system between the mean average planes of the interacting surfaces could prevent the Casimir force to induce stiction in actuating systems [30,35–37]. On the other hand, in direct bonding applications, the strong Casimir force between relatively smooth surfaces ($d_0 < 2$-$3$ nm) would be highly beneficial [16]. The aforementioned brought us to investigate also the effect of surface roughness from the different faces of SiC on the Casimir force, and compare the forces to those of thin film ruthenium (Ru) surfaces with comparable smoothness to that of SiC. The chosen Ru surfaces, which are relatively resistant to oxidation and are heavily used in multilayer coatings for mirrors in advanced lithography technologies [21], allow one to evaluate the relative effect of material conductivity changes at short interaction ranges, where the roughness effects are manifested as deviations from the Lifshitz theory predictions for flat surfaces.

## II. Optical characterization of SiC and Ru samples

Conductive nitrogen (N)-doped SiC with distinguished Si and C terminated faces were provided from Norstel AB (www.norstel.com) fabricated using the hot-wall technique. The Ru samples were obtained from the Industrial Focus Group XUV Optics, University of Twente, The Netherlands. Because the Casimir force is strongly influenced by the optical properties of the interacting bodies [38,39] we measured the optical properties of all the samples under study. Therefore, we performed Ellipsometry measurements in a wide range of frequencies using the J.A Woollam Co., Inc. ellipsometers VUV-VASE (0.5-9.34 eV) and IR-VASE (0.03-0.5 eV) at three different incident angles (55º, 65º, 75º) for SiC, and (70º, 75º, 80º) for Ru with respect to the sample surface. The optical data were analyzed as in [40], and subsequently the frequency dependent dielectric function $\varepsilon(\omega)$ was obtained. The latter allows calculation of the dielectric function $\varepsilon(i\zeta)$ at imaginary frequencies $\zeta$ (see Fig. 1 and Appendix), which is the necessary input for the Casimir force predictions via the Lifshitz theory.

Moreover, for the Casimir force calculations using the Lifshitz theory in the frequency range that is not covered by the experimental optical data, the Drude model was used to extrapolate at low frequencies (see Appendix) [41]. After fitting the optical data, the Drude parameters, namely, the plasma and relaxation frequencies ($\omega_p, \omega_\tau$) were (0.138 ± 0.008 eV, 0.074 ± 0.001 eV) for the SiC-Si face, (0.156 ± 0.008 eV, 0.078 ± 0.001 eV) for the SiC-C face, and (5.98 ± 0.07 eV, 0.09 ± 0.002 eV) for the Ru, respectively. From the plasma frequency $\omega_p$ we can estimate the concentration of conduction carriers $N_e = m^* \omega_p^2 / 4\pi e^2$ in SiC, if the effective carrier mass $m^*$ is close to the mass of the free electrons. The estimated $N_e$ for SiC-Si, SiC-C and Ru were ~1.4× $10^{19}$ cm$^{-3}$ (which is similar to the value for the SiC samples in our previous studies [28]), ~1.8× $10^{19}$ cm$^{-3}$ and ~2.6× $10^{22}$ cm$^{-3}$, respectively. The phonon polariton absorption peak for SiC, which is due to the absorption of infrared light by transverse optical phonon modes, was slightly higher for the SiC-Si face as compared to that of the SiC-C face. However, the difference between the dielectric function of the SiC-Si/C surfaces is negligible (see Fig. 1) implying that the calculated Casimir force via Lifshitz theory will not show any significant variation. This is because the penetration depth (skin effect) of the photons during ellipsometric measurements, which give the averaged value for $\varepsilon(\omega)$ within this depth, is not sensitive enough to capture the influence of the Si/C surface termination layers.

### III. Direct and inverse atomic force microscopy morphology measurements

The morphology of the surfaces of all samples was measured using a Bruker Multimode 8 atomic force microscope (AFM) operated in tapping mode to minimize the surface damage by the AFM tip. Figure 2 shows the AFM topography of the SiC-Si/C and Ru surfaces. The root-mean-square (rms) roughness was measured over a scan area of 1x1 μm$^2$, which is comparable to the effective Casimir force interaction area $\sim\pi dR$ in the sphere-plate geometry (for sphere radius R~10 μm and separations d<100 nm). The topography analysis yielded for the SiC-Si, SiC-C, and Ru surfaces $w_{SiC-Si} = 0.22$ nm, $w_{SiC-C} = 1.02$ nm, and $w_{Ru} = 0.45$ nm, respectively. Therefore, the surfaces of all samples are smooth enough (w ≤ 1 nm) to give only limited contribution to the separation upon contact between the sample and sphere surfaces, which is mainly limited by the roughness of the sphere.

Furthermore, in order to measure the Casimir force in the sphere-plane geometry, a borosilicate sphere with radius R=10.1 ± 0.6 µm was glued to the end of a tipless cantilever. The spheres were coated with Au using a Cressington 208 HR Sputter Coater. The sample stage in the coater can be rotated allowing coating of the spheres with homogenous thickness Au films which ensure electrical contact with the rest of the cantilever. The latter is necessary for the electrostatic calibration of the force measuring system (see Supplemental Material). The topography of the sphere is imaged using the so-called reverse AFM (see Fig. 2), where the sphere is scanned on top of a grid with inverted sharp tips (TGT1 grating from NT-MDT Spectrum Instruments, https://www.ntmdt-si.com/). This was performed in order to obtain topography information of the sphere surface as close as possible to the real interaction area between sphere-plate during the force measurements.

Indeed, as it is shown in Fig. 2, the sphere is considerably rougher than the planar SiC and Ru samples. Hence, the roughness of the sphere yields the dominant contribution to limit the force measurement at separations $d_0 \approx d_{0sph} + d_{0plt}$ [16] with $d_{0sph}$ and $d_{0plt}$ the maximum positive full widths due to the highest peaks in the height histograms (negative widths in the histograms correspond to valleys) (e.g. see Fig. 2 for the sphere) for both the sphere surfaces and planar samples, respectively. These estimations yielded a maximum distance upon contact $d_0$ for the Au/SiC-Si, Au/SiC-C, and Au/Ru interacting surfaces with values 30.8 $\pm$ 2 nm, 31.2 $\pm$ 2.1 nm and 35 $\pm$ 2.3 nm, respectively. However, because $d_{0sph} \gg d_{0plt}$, the actual separation distance is more likely to be $d_0 \approx d_{0sph} + w_{plt}$ with $w_{plt}$ the rms surface roughness of the sample surfaces. The latter yields for the actual distance upon contact $d_0$ between surfaces the values 30.2 nm, 31.0 nm and 30.5 nm for the Au/SiC-Si, Au/SiC-C, and Au/Ru systems, respectively.

### IV. Electrostatic and XPS analysis of material surfaces

The measurement of the electrostatic force between the sphere and the plate is crucial for the precise measurement of the actual Casimir force. This is because one obtains from these measurements the contact potential $V_o$ between the two interacting bodies, and more accurate values for the cantilever spring constant than those obtained from noise measurements (see supplemental Material) [30]. The knowledge of $V_o$ is necessary in order to apply potential compensation during the force measurements and minimize the electrostatic contributions on the measured force besides that of the genuine Casimir force. Hence, in order to obtain the contact potential, we applied various voltages V between the sphere and plate in the range [-3 V, 3 V]. The potential was applied to the sample with the tip grounded, while the tests with inverse polarity gave similar results. From the measurements of the electrostatic forces at larger separations ($\geq$ 100 nm), where the Casimir force plays a negligible role (by comparing to Casimir force predictions using Lifshitz theory for the size of the spheres used for the force measurements), we assigned the contact potential $V_o$ to the minimum of the cantilever deflection versus applied voltage V.

The contact potential $V_o$ (typically < 1 V) always exists between two dissimilar conductive materials (even between the same material prepared under different conditions), which are in thermal equilibrium, due to differences in work functions [31]. From the electrostatic force measurements in Fig. 3, we obtained a considerable difference between the measured contact potential $V_o$ for the Au/SiC-Si ($V_0^{Si}$ ~ -0.7 to -0.8 V) and Au/SiC-C ($V_0^{C}$ ~ -0.1 V) interacting surfaces. For $V_0^{Si}$ similar values were also found in our previous studies of Si-terminated (with comparable doping) SiC samples [28]. The measured contact potential for Au/Ru is $V_0^{Ru}$ = 0.05 V in our setup. The contact potential difference $\Delta V_o = V_0^{Si} - V_0^{C}$ (~0.6-0.7 V) between the Si/C-faces is significant even when taking into account that the C-face is relatively rougher than the Si face, and only chemically polished. However, besides trapped charges and any patch potentials (areas with different surface potential) caused by roughness, the electrostatic difference $\Delta V_o$ between the two SiC faces has to be attributed to other reasons.

Since the contact potential is linked to the work function difference between the two materials involved for the contact potential measurement and the work function consists of a bulk and surface part [42], the obtained contact potential difference between the Si- and C-face of SiC can be mainly explained by the difference of the status of the respective surface. In other words, because the bulk part of the work function is the same for the Si- and C-face the surface part of the work function must differ. According to the measured contact potential values, the surface part of the work function of the Si-terminated surface is by 0.6 – 0.7 V lower than the one of the C-terminated surface. This has to be related to differences in the chemical environment of the atoms present at the surface (and perhaps subsurface region) as well as the concentration of the types of atoms at the surface and subsurface region. Importantly, one has to consider that the above measurements were conducted under ambient conditions. This implies that oxidation of the surface needs to be taken into account when looking for an adequate explanation of the obtained data.

To obtain information on the types of different chemical species and their relative amounts present at the (sub)surface regions at the Si and C faces of SiC and with this to present an explanation of the observed contact potential differences, we conducted x-ray photoelectron spectroscopy (XPS) measurements (Figure 4, see also in Supplemental Material Fig. SM1 and Tables SM1 and SM2). The XPS data show that the chemical environment for C, N, O and Si is different on the Si- and C-faces [43]. Besides the expected higher amount of C and Si on the C and Si faces, respectively, the XPS results indicate that the amount of N (the dopant of the SiC samples) as well as O (due to oxidation at ambient conditions) is noticeably higher on the C face (see Supplemental Material Fig. SM1 and Tables SM1 and SM2). It should be noted that only a relative comparison of the amount of N and O on the two different sample surfaces is possible and that we did not perform an absolute quantification. The higher amount of nitrogen is based on the model that N prefers to occupy a C site because of the similarities in the covalent radii between the C and N atoms (covalent radius: Si: 0.110 nm, C: 0.077 nm, and N: 0.075 nm) [44]. Moreover, Si-N bonds are energetically more favorable than C-N ones so that N should rather incorporate in the C planes within SiC making more Si-N bonds compared to Si-C ones. The higher amount of oxygen means that the oxidation of the C face of 4H SiC proceeds much easier and faster compared to the Si face [33]. This can be on the one side explained by the difference in electronegativity of Si and C atoms when bound with O atoms (e.g. difference in Pauling electronegativity is $\Delta X_{CSi} = X_C - X_{Si} \approx 0.65$) [45]. On the other side, - and that is most likely the more relevant factor for the stronger oxidation of the C face - the increased presence of N on the C face will result in a higher amount of $NO_x$ compared to the Si face and thus, an overall higher amount of oxidation of the C face.

From previous investigations for adsorption of NO on a metal surface [46], it turned out that the work function increases compared to the bare metal surface. This is assigned to the dipole moment of NO (N being positively charged and O negatively) and the fact that NO binds with its N atom adsorbed on the metal surface. Transferring this knowledge to our system leads to the conclusion that a larger amount of $NO_x$ at the surface should also result in a higher work function, which is indeed what we obtained from the contact potential measurements.

The electrostatic findings for the contact potential of SiC generated further interest to perform a microscopic analysis of the SiC surfaces via Electric and Kelvin Probe Force Microscopy (EFM / KPFM) as shown in Fig. 5 [34]. Using EFM, one can measure electric field gradient distributions above the sample surface. It is also used to identify trapped charges and more importantly provide an electric polarization map of the sample surface. For EFM measurements the phase shift signal (~electric field gradient) is usually measured [35]. On the other hand, KPFM shows the local variation of the work function of materials, and it has been also previously used for this purpose in Casimir research [47]. However, KPFM images can change with adsorption layers like water, oxide layers, electrostatic charges, dopant concentration, and/or surface dipole moments [35]. Hence, we measured with KPFM the surface potential distribution over a scan area of $1\times1$ µm$^2$ for the three different samples (Figure 5a). For the SiC-C and Ru samples, the potential scale of the KPFM measurements did not match with the measured contact potential $V_0$ from the electrostatic calibration, while the SiC-Si face shows a relatively comparable scale variation with the measured $V_0^C$. Moreover, after performing EFM measurements, the SiC-Si face responded to the applied bias potential (Figure 5b). In fact, by changing the applied DC bias potential from 2 V to −0.5 V bright regions became darker and vice versa indicating the presence of surface charges. The EFM and KPFM results show that the surface of the SiC-Si face is electrically different from those of the SiC-C and Ru samples (see in Supplemental Material Fig. SM2) deviating from a metallic behavior, as it was also concluded from the obtained contact potentials via electrostatic calibration.

## V. Casimir force analysis

In order to calculate the force from experimentally obtained cantilever deflection data, the electrostatic calibration [30] of the cantilever-sphere spring constant was performed, which yielded k = 1.83 ± 0.02 N/m (see Supplemental Material) [36]. Moreover, the contribution of the separation dependent repulsive hydrodynamic drag force in the thin gap between sphere-plate surfaces is negligible. Indeed, this force is given by the expression [37] $F_h(z) = -(6\pi\mu R^2/z)(dz/dt)f^*$, where $\mu$ is the dynamic viscosity of the surrounding fluid (e.g., $\mu \approx 1.983 \times 10^{-5}$ kg/ms for air at T = 300 K), $dz/dt$ is the velocity of the sphere, and $f^*$ is the correction for the deviation from standard Reynolds flow due to fluid slip of the interacting surfaces. Assuming that no slip occurs ($f^* = 1$), then the piezo approach/retraction speed of dz/dt = 300 nm/s yields a repulsive hydrodynamic force $F_h \approx 3.2 \times 10^{-7}$ nN at short separations (e.g. $z \approx d_o \approx 35$ nm). The latter is negligible in comparison to Casimir forces of the order of pN to nN at the separations of interest ≤ 100 nm.

Furthermore, the Casimir force curves from both the theoretical calculation and the experimental data for the different samples are presented in Fig. 6. The relative thermal correction at T = 300 K for a separation below 100 nm can be neglected, so that we can use the convenient integral representation of the Lifshitz formula (see Appendix) to calculate the Casimir force [48]. Moreover, for completeness, we performed also force calculations via the Lifshitz theory using a variation of the dielectric function for the Au coating of the sphere within its maximum limits by considering the handbook data with the maximum plasma frequency of 9.0 eV. As it is shown (see in Supplemental Material Fig. SM3) the force difference for the Au-SiC system is negligible to play any role. One can also consider the plasma model to extrapolate at low frequencies (see Appendix) since this an unresolved issue for more than 15 years in the Casimir field, e.g., a signature of either an inconsistency in the Lifshitz theory or a contribution of electrostatic surface potentials [48, 4]. However, the results, at the short separations we probe (< 100 nm), will not change more than the variation we considered above for the Au coating of the spherical probe since at short separations (< 100 nm) the difference between Drude and Plasma models is not significant [49]. Furthermore, for a clear comparison between the experimental results and the Lifshitz theory calculation, we also calculated the relative force error $|F_{exp} - F_{Lif}|/F_{exp}$. The cantilever deflection was translated to a Casimir force using the electrostatically determined spring constant k = 1.83 N/m, and for the Lifshitz theory calculations we used the measured optical data of SiC-Si/C and Ru from Fig. 1, as well as those for Au from previous measurements [50].

The maximum absolute force difference is up to ~ 60% between the force measured at the closest distance $d_o$~30-35 nm, and the Lifshitz theory for Ru. The measured deviation is due to the sphere roughness contribution to the Casimir force at separations below 40 nm, where the high surface peaks lead to a rapid increase of the force, as it was observed in the past for the rough Au-Au systems [15, 20, 29]. Strong deviation was also observed for the SiC-C face, for which at the shortest separation the deviation due to roughness is ~40 %, while that for the SiC-Si face is about ~10-20 %. The roughness contribution appears to be of less importance as the conductivity of the probing sample reduces (since the sphere was the same for all surfaces), taking also into account that the SiC-Si/C and Ru surfaces have low rms surface roughness (w≤1 nm) over the effective

Casimir force interaction area of ~1x1 µm². Beyond 40 nm separations, the force data and Lifshitz theory data deviate only within a range of 10 %.

Beyond the separation regime, where surface roughness enhances the Casimir force (< 40 nm), one can describe the dependence of the Casimir force with an average power-law behavior $F_C \sim z^{-m}$ having an exponent m < 3 for the sphere-plate geometry. In fact, the obtained values for the exponents from the force data (and also from theory) were for the Ru, SiC-Si and SiC-C system 2.74 (theory: 2.53), 2.77 (theory: 2.70), and 2.65 (theory: 2.69), respectively. These values are in good agreement with the results of previous studies on Si-terminated SiC [28, 40] and other surfaces [51] that yielded m < 3 for the sphere-plate geometry over a diverse variety of interacting systems, and separation ranges < 1 µm. From the power-law behavior one can estimate the relative force error which occurs because of both the uncertainty in the actual surface separation due to the roughness contribution in $d_o$, and the error for the measured spring constant k. The estimated relative error for the Casimir force was calculated as $\Delta F_C/F_C \approx [(\Delta k/k)^2 + (m \, \Delta z/z)^2]^{1/2}$. The latter is roughly ~5 % for all three samples at the separation z = 100 nm and decreases with increasing separation for each sample. For the smoothest sample (SiC-Si face) at the shortest separation of z = 30.2 ± 1.8 nm, the relative error is $\Delta F_C/F_C$ ~15% (see also e.g. error bars in Fig. 6b). For the other surfaces, the relative estimated error at the shortest separations was also comparable. In any case, the estimated errors are significantly less than the force variation due to surface roughness in Fig. 6c, which can reach a level close to ~ 60 % for the most conductive surface at shortest separations of ~ 30 nm.

The comparison of the force data obtained with Lifshitz theory is only indicative due to the following limitations. First of all, the difference in charge carrier density for the two different faces of SiC, which was distinctly shown by electrostatic measurements was not translated into differences of the force calculated with the Lifshitz theory. This is because the ellipsometric measurement of the optical properties is not sensitive enough to variations of the dielectric permittivity of the different terminated surface layers. Similar interesting observations have been made in the past for other systems, where the surface layers were different from the bulk of the samples [52]. Second, comparing the experimental data with the theoretical prediction shows that there is a considerable deviation from the Lifshitz theory for flat surfaces at short separations (< 40 nm). The reason for this discrepancy is due to the considerable effect of roughness on the Casimir force due to the high surface peaks [20, 29], which becomes more pronounced for material systems with increasing surface conductivity (Fig. 6c). The surface roughness also has a strong effect in determining the actual separation distance between the interacting bodies [16, 29]. The experimental data for Ru appear to be noisier at larger separations as compared to both SiC samples. The latter can be attributed to possible degradation of the probe since all force measurements for the three samples were performed by one individual probe.

## VI. Conclusions

In summary, the comprehensive knowledge of the Casimir force as well as the associated electrostatic characteristics from differently terminated conductive SiC and Ru surfaces can be essential in diverse micro/nanotechologies for operation in harsh environments and the design of advanced coatings in optics. Therefore, we performed Casimir force measurements between the same Au-coated microsphere and N doped SiC with both Si-terminated and C-terminated faces and compared the results to relatively flat and metallic Ru surfaces. Electrostatic calibration measurements showed a ~0.6-0.7 V difference in the contact potential $V_o$ between SiC-Au for the two faces of SiC. We attributed this to the higher incorporation of N and subsequently the formation of $NO_x$ on the C face which was confirmed by XPS measurements. On the other hand, the measured optical properties of the SiC-Si/C terminated surfaces with ellipsometry did not show any substantial differences indicating that the effective depth of the different Si/C surface layers is significantly smaller than the skin depth prohibiting as a result any substantial differences to appear in force calculations via Lifshitz theory.

However, the Casimir force measurement after minimization of electrostatics contributions showed differences between the Si/C faces, while the comparison with the Lifshitz theory calculations show better agreement for the SiC-Si face. In addition, short-range force measurements at separations less than 40 nm were mainly limited by the surface roughness of the sphere. In fact, the comparison of the Casimir forces below 40 nm between the SiC-Si/C and Ru surfaces indicates that the short-range roughness effects increase in magnitude with increasing metallic behavior of the sample surface. Although comparisons with Lifshitz theory are limited for the samples under considerations, our results indicate that surface layers and surface roughness must be taken into consideration in Casimir and electrostatic force measurements as dimensions and surface separations decrease from the micro to the nano regimes.

**Acknowledgements**

We kindly thank Prof. F. Bijkerk for providing the Ru sample, and Gert ten Brink for the technical support. This research was funded by the Netherlands Organization for Scientific Research (NWO) under Grant number 16PR3236.

**Appendix: Lifshitz theory and extrapolation of optical data**

The Casimir force $F_{Cas}(d)$ in Eq.(2) is given by [6]

$$F_{Cas}(d) = \frac{k_B T}{\pi} {\sum_{l=0}}' \sum_{\nu=TE,TM} \int_0^\infty dk_\perp \, k_\perp \, k_0 \frac{r_\nu^{(1)} r_\nu^{(2)} \exp(-2k_0 d)}{1 - r_\nu^{(1)} r_\nu^{(2)} \exp(-2k_0 d)}. \tag{A1}$$

The prime in the first summation indicates that the term corresponding to $l = 0$ should be multiplied with a factor 1/2. The Fresnel reflection coefficients are given by $r_{TE}^{(i)} = (k_0 - k_i)/(k_0 + k_i)$ and $r_{TM}^{(i)} = (\varepsilon_I k_0 - \varepsilon_0 k_i)/(\varepsilon_I k_0 + \varepsilon_0 k_i)$ for the transverse electric (TE) and magnetic (TM) field polarizations, respectively. $k_i = \sqrt{\varepsilon_I(i\xi_l) + k_\perp^2}$ ($i = 0,1,2$) represents the out-off plane wave vector in the gap between the interacting plates ($k_0$) and in each of the interacting plates ($k_{i=(1,2)}$). $k_\perp$ is the in-plane wave vector.

Furthermore, $\varepsilon(i\xi)$ is the dielectric function evaluated at imaginary frequencies, which is necessary for calculating the Casimir force between real materials using Lifshitz theory. Applying the Kramers-Kronig relation, $\varepsilon(i\xi)$ is given by [12]

$$\varepsilon(i\xi) = 1 + \frac{2}{\pi} \int_0^\infty \frac{\omega \, \varepsilon''(\omega)}{\omega^2 + \xi^2} \, d\omega. \tag{A2}$$

For the calculation of the integral in Eq. (A2) one needs the measured data for the imaginary part of the frequency dependent dielectric function $\varepsilon''(\omega)$. The experimental data for the imaginary part of the dielectric function cover only a limited range of frequencies $\omega_1 \, (= 0.03 \text{ ev}) < \omega < \omega_2 \, (= 8.9 \text{ ev})$. Therefore, for the low optical frequencies ($\omega < \omega_1$) we extrapolated using the Drude model [12, 35]

$$\varepsilon''_L(\omega) = \frac{\omega_p^2 \omega_\tau}{\omega(\omega^2 + \omega_\tau^2)}, \tag{A3}$$

$\omega_p$ is the Plasma frequency, and $\omega_\tau$ is the relaxation frequency. For the higher optical frequencies ($\omega > \omega_2$) we extrapolated using the expression [12, 31, 35]

$$\varepsilon''_H(\omega) = \frac{A}{\omega^3} \tag{A4}$$

Using Eq. (A2)-(A4), $\varepsilon(i\xi)$ in terms of the Drude model is given by [31, 35]

$$\varepsilon(i\xi)_D = 1 + \frac{2}{\pi} + \int_{\omega_1}^{\omega_2} \frac{\omega \, \varepsilon''_{exp}(\omega)}{\omega^2 + \xi^2} \, d\omega + \Delta_L \varepsilon(i\xi) + \Delta_H \varepsilon(i\xi) \tag{A5}$$

with

$$\Delta_L \varepsilon(i\xi) = \frac{2}{\pi} \int_0^{\omega_1} \frac{\omega \, \varepsilon''_L(\omega)}{\omega^2 + \xi^2} \, d\omega = \frac{2\omega_p^2 \omega_\tau}{\pi(\xi^2 - \omega_\tau^2)} \left[ \frac{\arctan\left(\frac{\omega_1}{\omega_\tau}\right)}{\omega_\tau} - \frac{\arctan\left(\frac{\omega_1}{\xi}\right)}{\xi} \right], \tag{A6}$$

and

$$\Delta_H \varepsilon(i\xi) = \frac{2}{\pi} \int_{\omega_2}^\infty \frac{\omega \, \varepsilon''_H(\omega)}{\omega^2 + \xi^2} \, d\omega = \frac{2\omega_2^3 \, \varepsilon''(\omega_2)}{\pi \xi^2} \left[ \frac{1}{\omega_2} - \frac{\frac{\pi}{2} - \arctan\left(\frac{\omega_2}{\xi}\right)}{\xi} \right]. \tag{A7}$$

If extrapolation at low frequencies is performed with the plasma model, then one must replace the term $\Delta_L \varepsilon(i\xi)$ in Eq.( A6) with $\omega_p^2/\xi^2$. Therefore, for the Plasma model $\varepsilon(i\xi)$ is given by [39]

$$\varepsilon(i\xi)_P = 1 + \frac{2}{\pi}\int_{\omega_1}^{\omega_2}\frac{\omega\varepsilon''_{exp}(\omega)}{\omega^2+\xi^2}d\omega + \frac{\omega_p^2}{\xi^2} + \Delta_H\varepsilon(i\xi). \tag{A8}$$

# Figure Captions

**Figure 1** Dielectric function of SiC-Si, SiC-C and Ru at imaginary frequencies, which were calculated using the Drude model to extrapolate at low frequencies. The zoomed-in area shows that there is a small difference in the calculated dielectric function for the SiC-Si/C terminated layers. The inset shows the imaginary part of the frequency dependent dielectric function.

**Figure 2** The topography (3D height images) of (a) Ruthenium, (b) SiC-C face, (c) SiC-Si face and (d) the Au coated sphere. The inverse AFM image of the sphere shows a repeated pattern since the same contact area of the sphere is scanned by multiple sharp tips. (e) Height distribution of the three samples which yields from its positive width the maximum contribution $d_{o,plt}$ to the separation upon contact. The C face is only chemically polished and contains several scratches. (f) Height distribution of the sphere as obtained by inverse AFM.

**Figure 3** Cantilever deflection versus applied voltage in the range [-3V, 3V]. The contact potential $V_o$ is assigned at the minimum of the cantilever deflection: (a) Au/SiC-C, and (b) Au/SiC-Si systems. The different curves correspond to different sphere-plate separations in the range $\geq 100$ nm where the Casimir force has negligible contribution. The lines are only a guide for the eye.

**Figure 4** Comparison of XPS spectra for the Si- and C-face of SiC. (a-c) Si 2p, (d-f) C 1s, (g-i) N 1s and (j-l) O 1s core level spectra. The peak assignment is detailed in Table SM2 in the Supplemental material.

**Figure 5** (a) KPFM results (Left: Height topography, and Right: Potential) of the SiC-Si face. (b) The EFM image of SiC-Si face with an applied potential of 2 volts (left) and -0.5 volts (right) between the sample and the tip.

**Figure 6** The experimental Casimir force data together with the Lifshitz theory calculation in the separation 30-100 nm. (a) Ru, (b) SiC-Si/C faces. (c) Relative force error comparing the experimental Casimir force data with Lifshitz theory calculations using the measured optical data for the three samples shown in Fig.1.

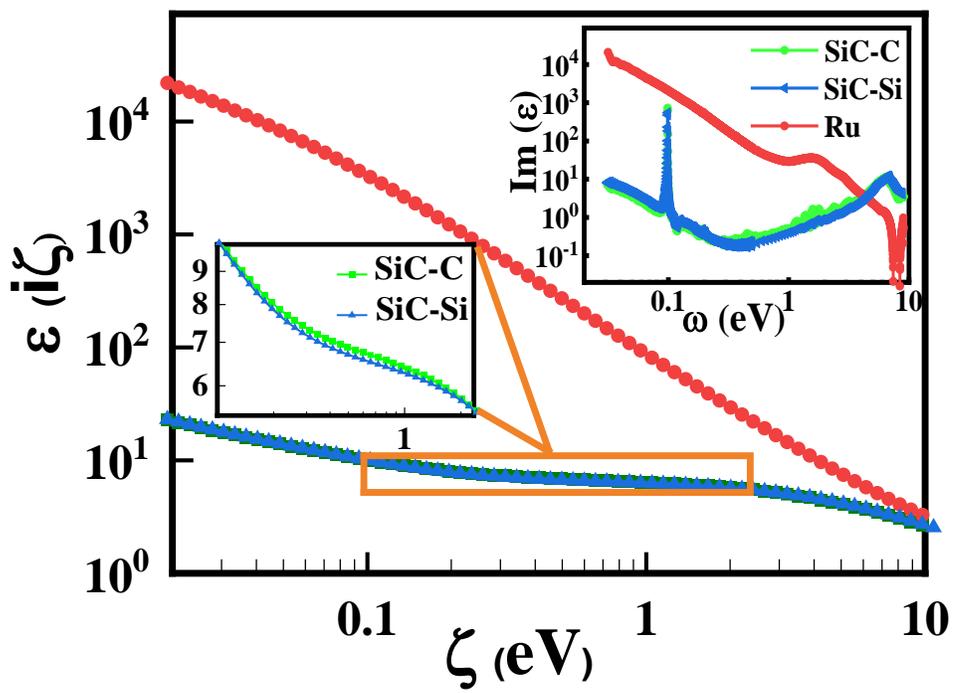

**Figure 1**

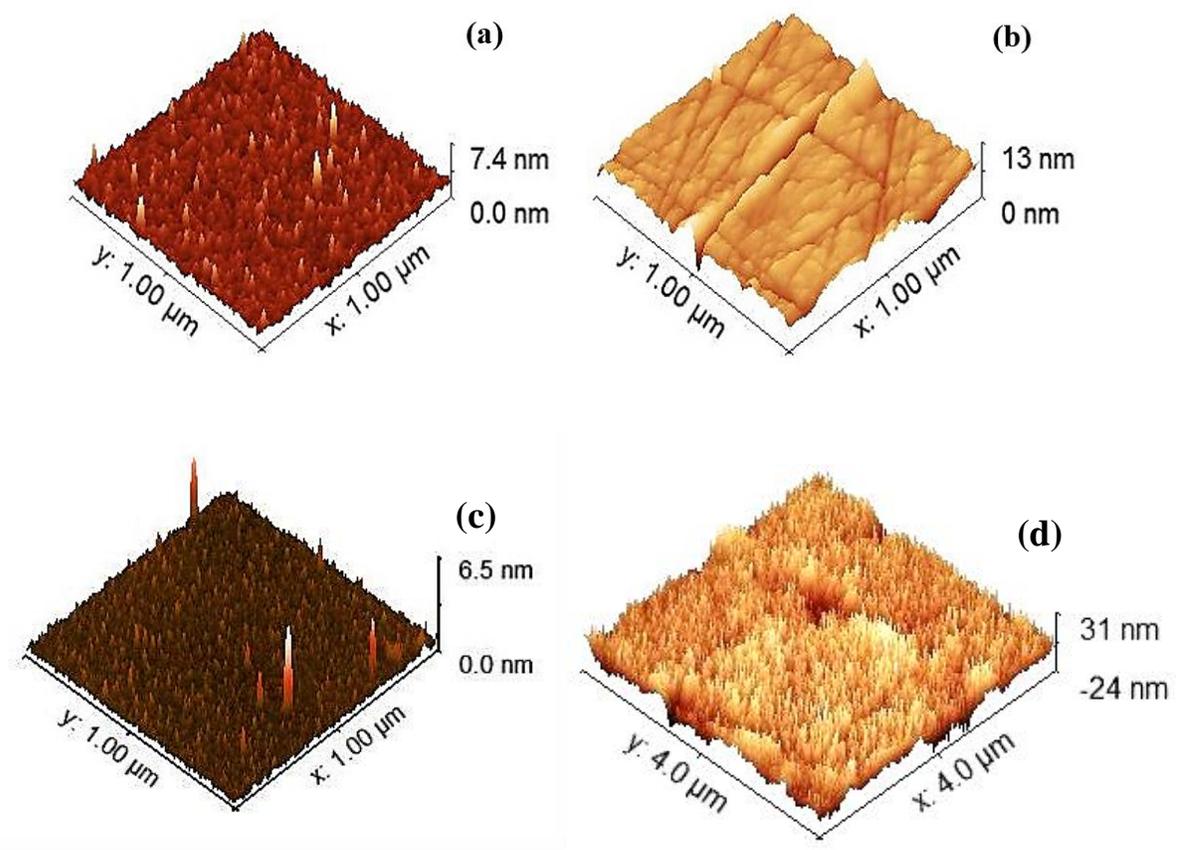

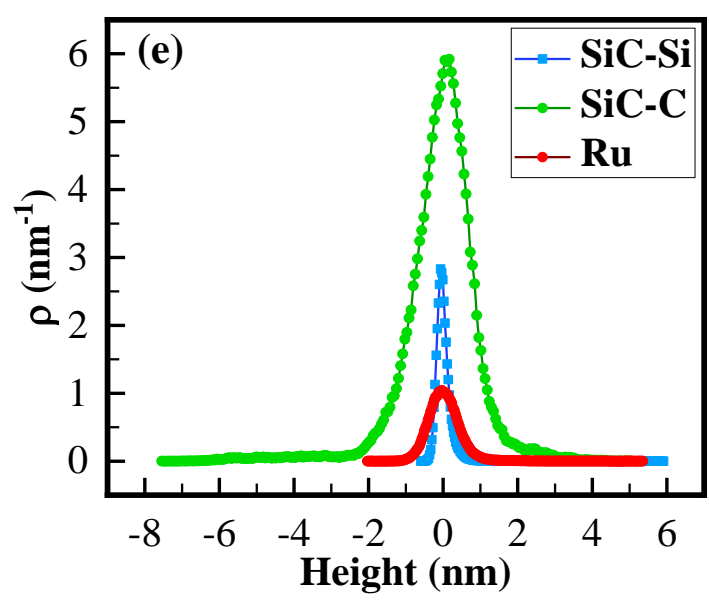

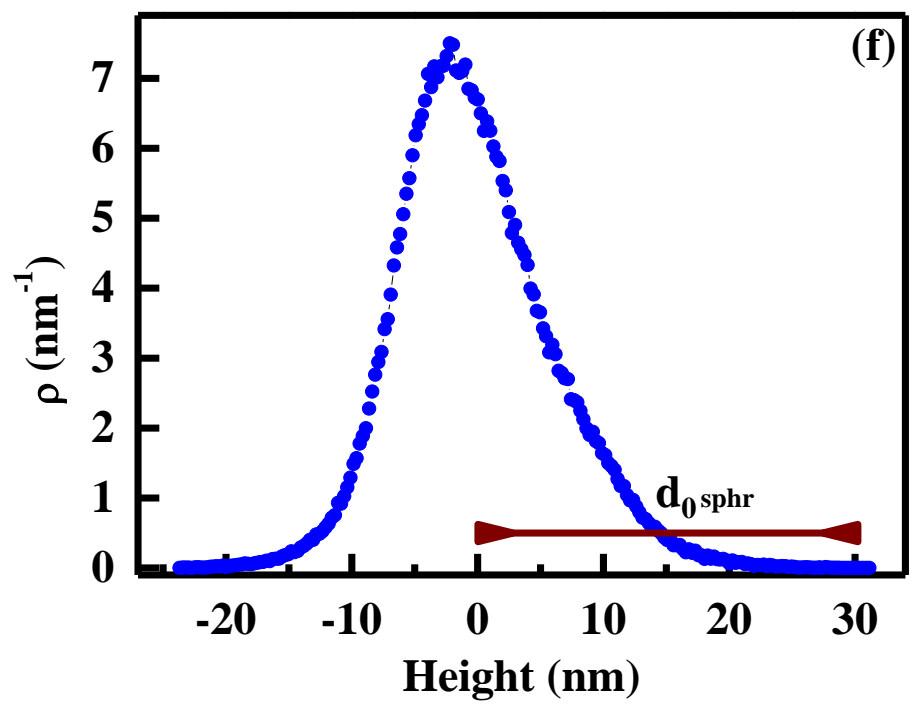

Figure 2

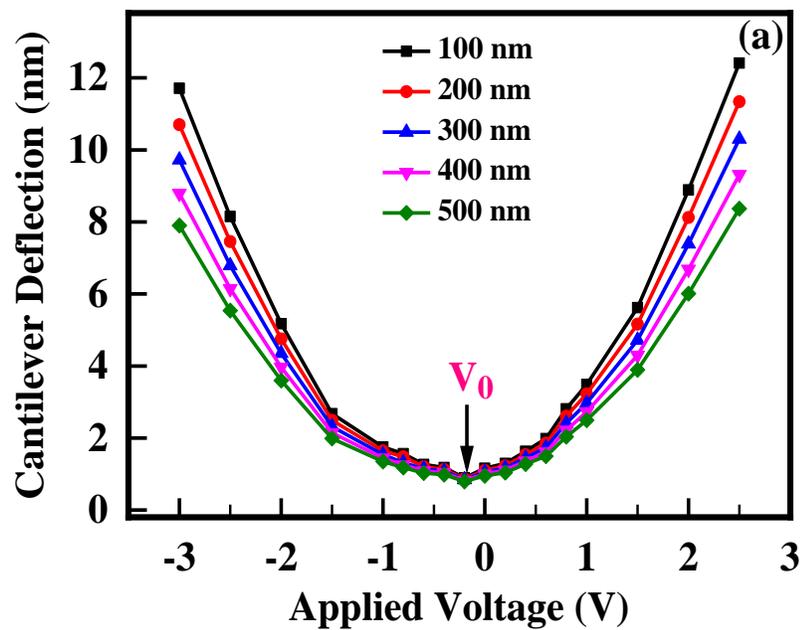
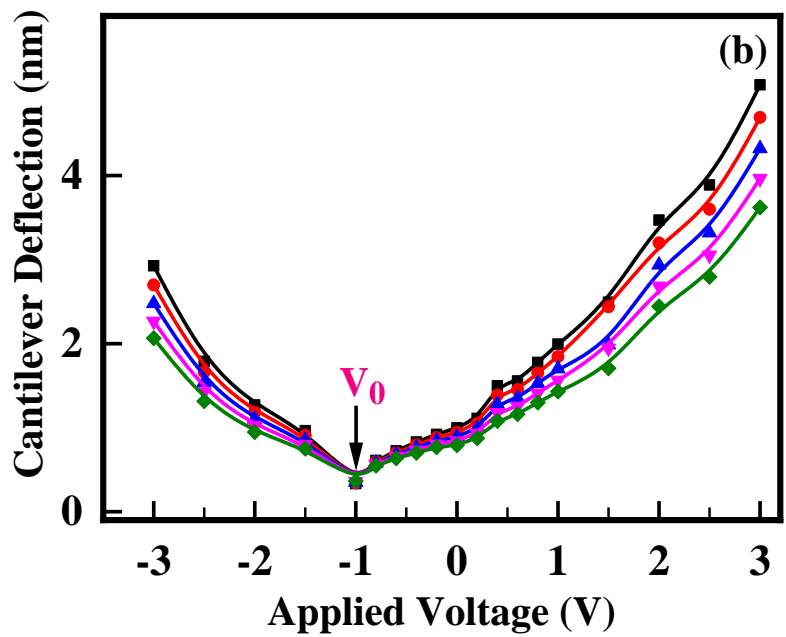

**Figure 3**

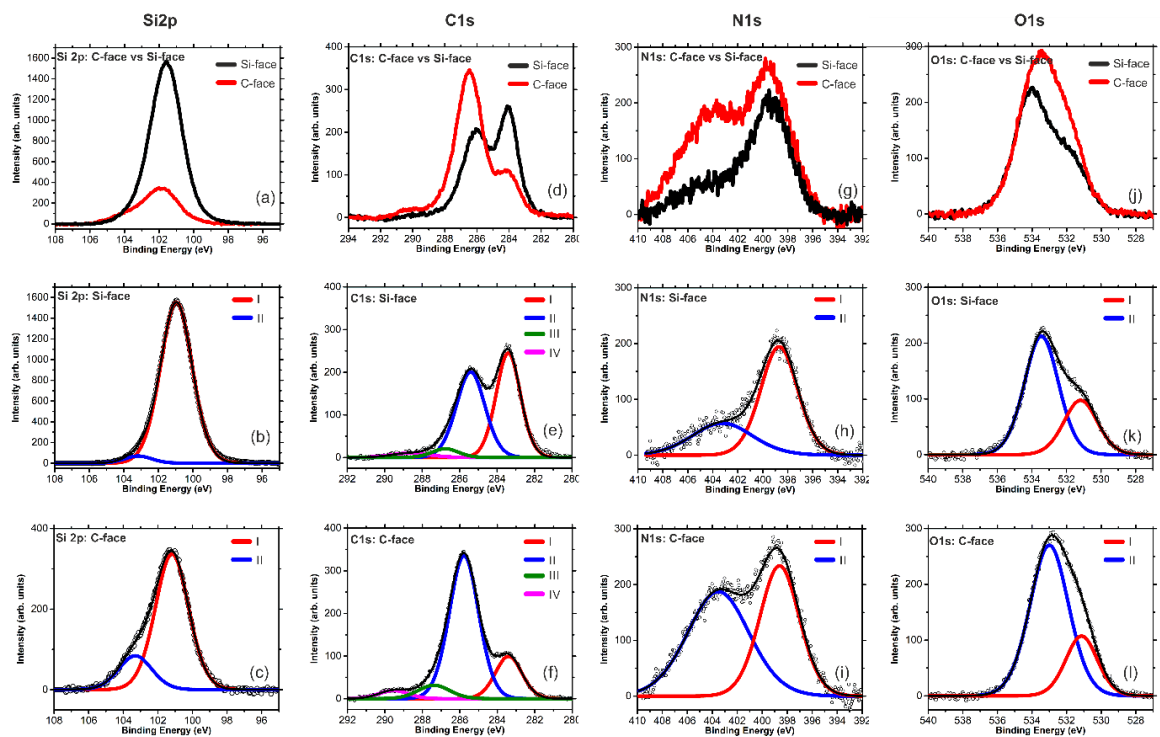

**Figure 4**

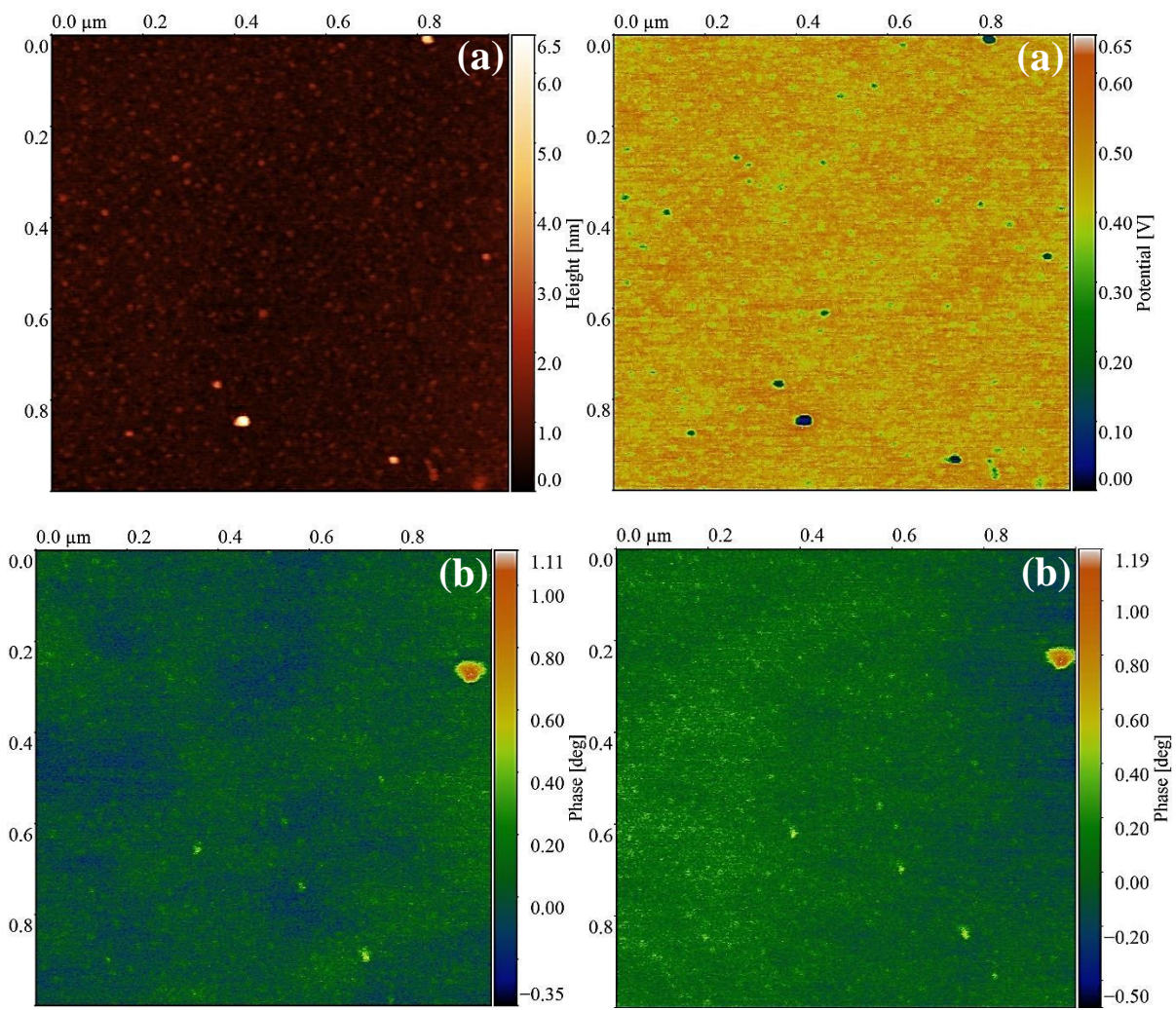

**Figure 5**

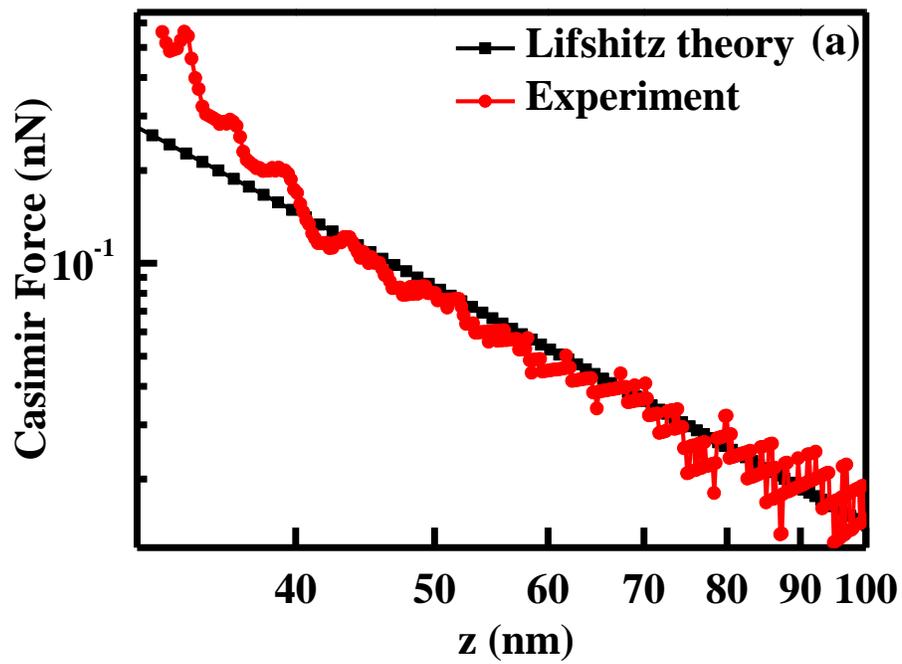

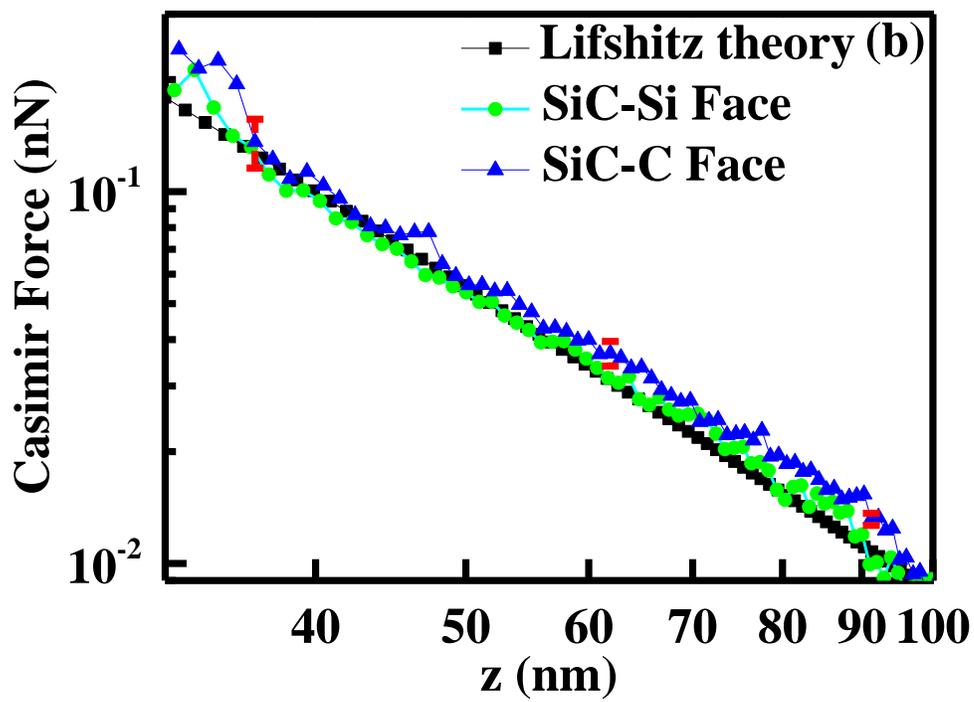

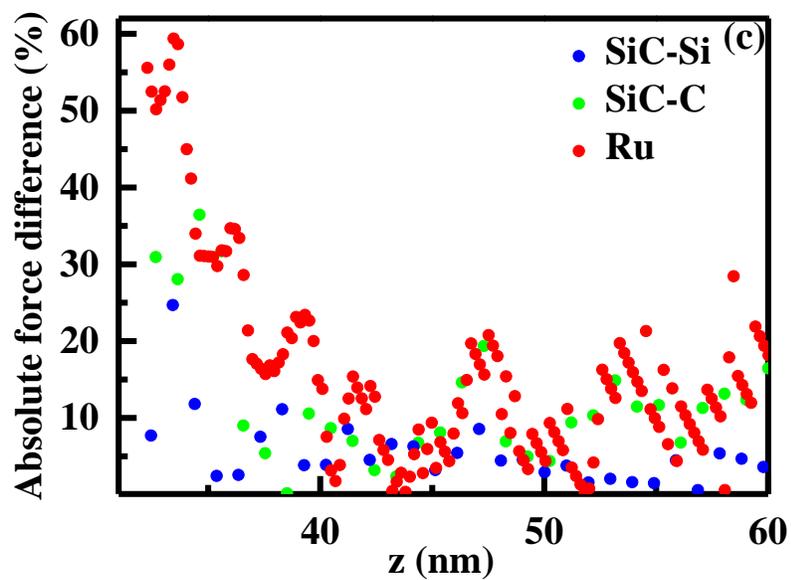

**Figure 6**